\documentclass[aps,prl,twocolumn,groupedaddress,showpacs]{revtex4}

\usepackage{graphicx}
\usepackage{amsmath}

\begin{document}

\title{Spin Accumulation in Quantum Wires with Strong Rashba
Spin-Orbit Coupling}

\author{M. Governale}
\author{U. Z\"ulicke}

\affiliation{Institut f\"ur Theoretische Festk\"orperphysik,
Universit\"at Karlsruhe, D-76128 Karlsruhe, Germany}

\date{\today}

\begin{abstract}
We present analytical and numerical results for the effect of
Rashba spin-orbit coupling on band structure, transport, and
interaction effects in quantum wires when the spin precession
length is comparable to the wire width.
The situation with only the lowest spin-split subbands occupied is
particularly interesting because electrons close to Fermi points
of the same chirality can have approximately parallel spins. We
discuss consequences for spin-dependent transport and effective
Tomonaga--Luttinger descriptions of interactions in the quantum
wire.
\end{abstract}


\maketitle

Spin-dependent transport phenomena are currently attracting a lot
of interest because of their potential for future electronic
device applications\cite{wolf:sc:01}. Basic design proposals for
spin-controlled field-effect switches\cite{spinfet,nitta:apl:99}
use the fact that electron waves with opposite spin aquire
different phase factors during their propagation in the presence
of Rashba spin-orbit coupling\cite{rashba} (RSOC). The latter
arises due to structural inversion asymmetry in quantum
heterostructures\cite{roess:prl:88,wink:prb:00} where
two-dimensional (2D) electron systems are realized. The
single-electron Hamiltonian is then of the form\cite{byra:jpc:84}
$H_{\text{2D}} = H_0 + H_{\text{so}}$ where
\begin{subequations}
\begin{eqnarray}
H_0 &=& \frac{1}{2 m}\left(p_x^2 + p_y^2\right) \quad , \\
H_{\text{so}} &=& \frac{\hbar k_{\text{so}}}{m}\left( \sigma_x\,
p_y - \sigma_y\, p_x\right)\quad ,
\end{eqnarray}
\end{subequations}
with $m$ denoting the effective electron
mass\footnote{Band nonparabolicity typically plays no r\^ole
in the low-density quantum wires considered here.}.
Possibility to tune the strength of RSOC, measured here in terms
of the characteristic wave vector $k_{\text{so}}$, by external
gate voltages has been demonstrated
experimentally\cite{nitta:prl:97,schaep:jap:98:sh,grund:prl:00}.
As a manifestation of broken spin-rotational invariance,
eigenstates of $H_{\text{2D}}$ which are labeled by a 2D wave
vector $\vec k$ have their spin pointing in the direction
perpendicular to $\vec k$. Hence, no common spin quantization axis
can be defined for eigenstates when spin-orbit coupling is
present. Confining the 2D electrons further to form a quantum
wire, one might naively expect to again be able to define a global
spin quantization axis, as the propagation direction of electrons
in a one-dimensional (1D) system is fixed. However, this turns out
to be correct only for a truly 1D electron system with vanishing
width. In real quantum wires, such a situation is approximately
realized when the spin-precession length\cite{spinfet} $\pi/
k_{\text{so}}$ is much larger than wire width. Another way to
formulate this condition is to say that the characteristic energy
scale $\Delta_{\text{so}}=\hbar^2 k_{\text{so}}^2/2 m$ for RSOC is
small compared to the energy spacing of 1D subbands. For a quantum
wire defined by a parabolic confining potential, e.g., 
\begin{equation}\label{confine}
V(x) = \frac{m}{2}\,\omega^2\, x^2\quad ,
\end{equation}
the latter would be $\hbar\omega$. When spin-orbit coupling is not
small (i.e., when $\Delta_{\text{so}}\sim\hbar\omega$ for the case
of parabolic confinement), hybridization of 1D subbands for
opposite spins becomes important, resulting in the deformation of
electronic dispersion relations\cite{moroz:prb:99}. The effect of
this deformation on transport properties has been the subject of
recent investigation\cite{moroz:prb:99}, e.g., with respect to
implications for the modulation of spin-polarized conductances as
a function of RSOC strength\cite{mire:prb:01} which is the
principle of operation for spin-controlled field-effect
devices\cite{spinfet,nitta:apl:99}.

Here we present results for the detailed spin structure of
electron states in a quantum wire, defined by the parabolic
confining potential $V(x)$ given in Eq.~(\ref{confine}), with
strong RSOC present. Contrary to previous\cite{moroz:prl:00}
assumptions that were uncritically adopted
in the recent literature\cite{egg:epl:01}, we find that electrons
with large wave vectors in the lowest spin-split subbands have
essentially parallel spin. The spin state that right-moving
electrons converge toward is opposite to that for left-movers.
This counterintuitive result will be explained qualitatively in
the following paragraph, before presenting analytical and
numerical results for electronic dispersion curves and spin
structure of eigenstates. A texture-like variation of spin density
{\em across\/} the wire is identified. We then apply the
Landauer-B\"uttiker formalism\cite{landauer2,butt:ibm:88} to
discuss spin--dependent transport in hybrid systems of a wire with
RSOC attached to leads where $k_{\text{so}}=0$. Current turns out
to be spin--polarized in the wire but unpolarized in the leads. We
elucidate the peculiar current conversion at wire--lead interfaces
that sustains this novel type of spin accumulation
in the wire. Finally, consequences for the low-energy description
of interacting wires in terms of Tomonaga-Luttinger-type models
are discussed.

\begin{figure}
\includegraphics[width=2.5in]{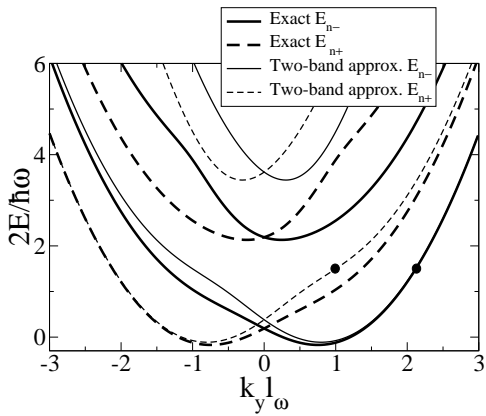}
\caption{Lowest and first excited spin-split subbands of a quantum
wire, defined by a parabolic confining potential with oscillator
length $l_\omega$ in a 2D electron system, with strong Rashba
spin-orbit coupling such that $k_{\text{so}}\, l_\omega=0.9$.
Thick curves are results of the exact numerical calculation, while
thin curves are obtained using the approximate two-band model
which includes only spin-orbit induced mixing of the lowest two
parabolic subbands. Evidently, this approximation gives reasonable
results for the lowest spin-split subband, even in the present
case of a rather large spin-orbit coupling
strength.\label{bandstruct}}
\end{figure}

We start by considering basic features for eigenstates of the
Hamiltonian $H_{\text{1D}}=H_{\text{2D}}+V(x)$ which are 1D plane
waves in the $y$ coordinate with wave number $k_y$ but bound in
$x$ direction. At finite $k_{\text{so}}$, spin degeneracy is
preserved only for eigenstates with $k_y=0$; their energies are
the shifted harmonic-oscillator levels $E_n^{(0)}=\frac{\hbar
\omega}{2}(2n+1)-\Delta_{\text{so}}$. This result is exact. To
characterize states with finite $k_y$, we rewrite $H_{\text{1D}}=
H_{\text{pb}} + H_{\text{mix}}$ where
\begin{equation}
H_{\text{pb}} = \frac{p_x^2}{2 m} + \frac{m\omega^2 x^2}{2}+
\frac{\hbar^2 k_y^2}{2 m} + \frac{\hbar^2 k_{\text{so}}k_y}{m}\,
\sigma_x \, ,
\end{equation}
and $H_{\text{mix}} = -\hbar k_{\text{so}}\sigma_y p_x/m$.
Straightforward calculation yields eigenstates of $H_{\text{pb}}$
which are also eigenstates of $\sigma_x$ with eigenvalue $\sigma=
\pm 1$ and have energies $E_{n\sigma}^{\text{(pb)}}(k_y) = \frac
{\hbar\omega}{2}(2 n+1)+\frac{\hbar^2}{2 m}\left(k_y + \sigma
k_{\text{so}}\right)^2 - \Delta_{\text{so}}$. The term
$H_{\text{mix}}$ induces mixing between the shifted parabolic
subbands $E_{n\sigma}^{\text{(pb)}}(k_y)$. To lowest order in
perturbation theory, it results in a uniform shift of
eigenenergies by $-\Delta_{\text{so}}$ and a small deviation of
spin quantization in $x$ direction\cite{haeus:prb-rc:01}. Hence,
for $\Delta_{\text{so}}\ll\hbar\omega$, eigenstates of
$H_{\text{1D}}$ have energies $E_{n\sigma}^{\text{(pb)}}(k_y)-
\Delta_{\text{so}}$ and are, to a good approximation,
eigenstates of $\sigma_x$. When $\Delta_{\text{so}}$ becomes
comparable to the subband splitting, anticrossings occur between
neighboring subbands with {\em opposite\/} spin index
$\sigma$. As a result, no common spin-quantization axis can be
defined anymore for eigenstates within any subband. Far enough
from anticrossings, eigenstates of $H_{\text{so}}$ will
essentially be eigenstates of $H_{\text{pb}}$. In particular,
their spins will be approximately aligned in $x$ direction. In the
lowest two subbands, right-movers with wave vectors larger than
that of the anticrossing point can then have approximately
parallel spin. The same is true for left-movers whose asymptotic
spin direction is opposite to that of right-movers.

In Fig.~\ref{bandstruct}, we show as thick lines numerically
calculated spectra of $H_{\text{1D}}$ for a large value of
spin-orbit coupling. Deviation from parabolicity is clearly
visible. Interestingly, it is possible to obtain a good
quantitative description of the lowest spin-split subband by
diagonalizing $H_{\text{1D}}$ in a truncated Hilbert space which
is spanned by the lowest and first-excited spin-degenerate
parabolic subbands of the Hamiltonian $H_0+V(x)$. We call this the
{\em two-band\/} model and find an approximate expression for the
dispersion of the lowest spin-split subband,
\begin{equation}\label{twobandlow}
\frac{2 E_{0\gamma}^{(\text{2b})}}{\hbar\omega}=2+(k_y l_\omega)^2
-\sqrt{(1-\gamma 2 k_{\text{so}} k_y l_\omega^2)^2+2(k_{\text{so}}
l_\omega)^2} \quad ,
\end{equation}
where $l_\omega=\sqrt{\hbar/m\omega}$ is the oscillator
length of the parabolic confinement, and $\gamma=\pm$ a
subband index that does {\em not\/} have the meaning of a spin
quantization number. We show Eq.~(\ref{twobandlow}) and the
corresponding result for the first excited subband as thin lines
in Fig.~\ref{bandstruct}. It is seen that the two-band
model is quite adequate for the lowest subbands, even for rather
strong spin-orbit coupling.

\begin{figure}
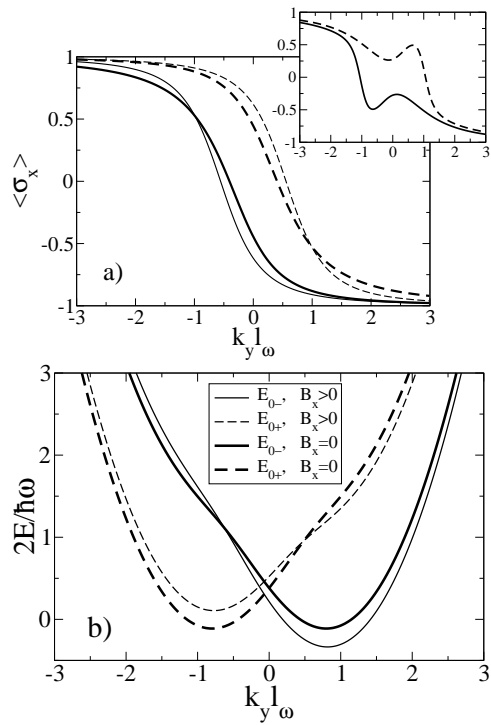

\includegraphics[width=2.5in]{avgboth.eps}
\includegraphics[width=2.5in]{bandb.eps}
\caption{Spin structure of electron states in a quantum wire with
strong spin-orbit coupling. Panel~a): Expectation value of spin
projection onto the $x$ direction for electron states obtained in
Fig.~\ref{bandstruct}. Results of exact numerical calculation for
the lowest spin-split subbands (main figure) and first excited
spin-split subbands (inset) are given by thick curves. The
effective two-band model reasonably approximates the behavior of
the lowest subband (thin lines in the main figure). Right-moving
electrons with large wave vectors asymptotically have parallel
spin which is opposite to that of left-movers. The same can be
observed in panel~b) where the spectrum in a finite magnetic field
$B$ pointing in $x$ direction is compared with that in zero field.
Here dispersion curves are calculated within the two-band model,
for Zeeman energy $g\mu_{\text{B}}B=0.25\hbar\omega$ (thin lines)
and in zero magnetic field (thick lines).\label{avspin}}
\end{figure}

\begin{figure}
\includegraphics[width=2.5in]{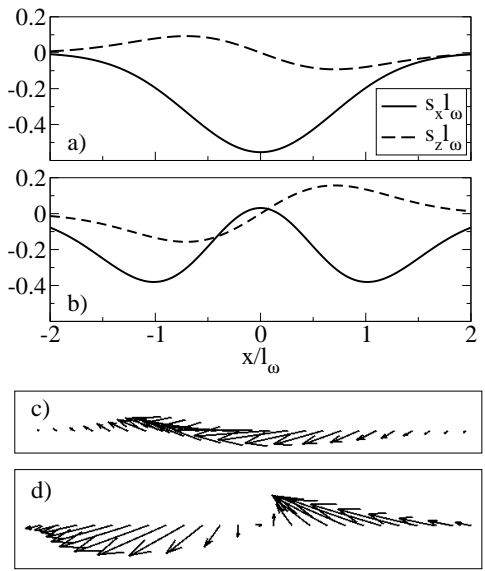}
\caption{Texture-like structure of spin density across the quantum
wire, calculated within the two-band model for states indicated by
black dots in Fig.~\ref{bandstruct}, which have energy $0.75\,
\hbar\omega$. a)~Spatial variation of nonzero components of spin
density for the state with larger wave vector. b)~Same for the
other state. c)~Visualization of spin texture for the same state
as a). Arrow length is proportional to spin density. d)~Spin
texture visualized for the same state as considered in b).
\label{spintex}}
\end{figure}

Results shown in Fig.~\ref{avspin} confirm conclusions reached in
our previous discussion of the spin structure of electron
eigenstates with RSOC present. Panel~a) shows the expectation
value of spin component in $x$ direction for eigenstates of
$H_{\text{1D}}$ in the lowest and first excited spin-split
subbands for the same value of $k_{\text{so}}$ used in
Fig.~\ref{bandstruct}. Data in Figs.~\ref{bandstruct} and
\ref{avspin} for the same subband are indicated by the same line
type. For the lowest subbands, we also give, as thin lines,
results obtained analytically within the two-band model. It is
clearly seen that spins of eigenstates with large absolute value
of wave number are approximately quantized in $x$ direction, with
spin direction of left-movers being opposite to that of
right-movers\footnote{This result contradicts the spin labeling
of subbands adopted in Refs.~\cite{moroz:prl:00,egg:epl:01}.}.
This fact is underscored by the properties of the energy spectrum
in a finite magnetic field $B$ in $x$ direction which is shown in
panel~b). Clearly, Zeeman shift of states at large positive wave
number is opposite to that for states with large negative wave
number. Shown as thin lines in the main figure of panel~a) are
curves obtained analytically within the two-band model which
yields again reliable results for the lowest subbands. We
therefore use it to calculate the variation of spin density $\vec
s(x)=\Phi^\dagger(x)\vec\sigma\Phi(x)$ {\em across\/} the wire.
[The spinor $\Phi(x)$ denotes the transverse part of an
eigenfunction of $H_{\text{1D}}$ which, in the presence of
spin-orbit coupling, depends on wave vector.] It turns out that
the density $s_y(x)$ of spin component parallel to the wire
vanishes identically. Hence, only the $x$ and $z$ components of
spin density are shown in Fig.~\ref{spintex}, displaying an
interesting texture-like variation with coordinate $x$ whose
structure reflects the mixing between subbands due to
$H_{\text{mix}}$. Note that the {\em expectation value\/} for the
$z$ component of spin vanishes for eigenstates of $H_{\text{1D}}$.

\begin{figure}
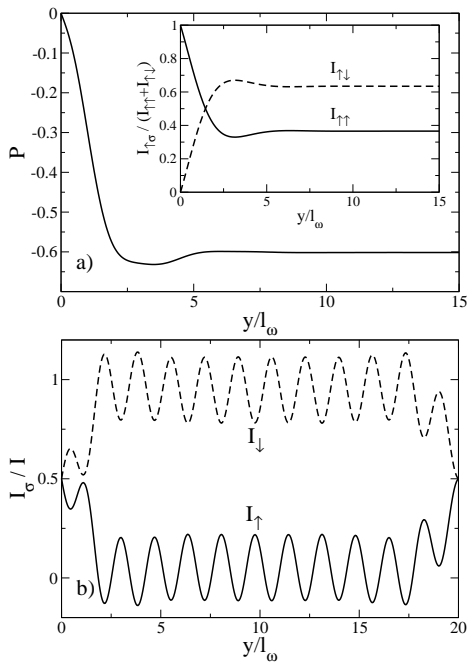

\includegraphics[width=2.4in]{1interface.eps}
\includegraphics[width=2.4in]{2interfaces.eps}
\caption{Transport in hybrid systems of a wire with strong Rashba
spin-orbit coupling and ideal leads, calculated exactly using the
Landauer--B\"uttiker formalism within the two--band model
(inclusion of higher subbands leads only to small quantitative
changes). Panel~a) shows the spatial
variation of current polarization in a semi-infinite wire ($y>0$)
attached to an ideal lead ($y<0$). Conversion of incident spin-up
current is illustrated in the inset. Here $I_{\sigma\uparrow}$ denotes
the spin--$\sigma$ current in the wire when spin--$\uparrow$ current is
injected from the lead.
A finite spin polarization
exists also in a finite wire with to semi-infinite leads attached
[panel~b)]. Here quantum interference gives rise to additional
oscillatory structure. Parameters used in the calculation are
$E_{\text{F}}=1\,\hbar\omega$ and $k_{\text{so}}=0.9\,
l_\omega^{-1}$ \label{convert}}
\end{figure}

From the above it has become clear that, in general, spin
quantum number is {\em not\/} an appropriate way to characterize
electron states in a quantum wire with strong RSOC. Only states
with wave number $k_y$ far enough from anticrossing points will
asymptotically have their spin quantized in $x$ direction. From
considering Figs.~\ref{bandstruct} and \ref{avspin}, the following
special situation can be envisioned which has rather
counterintuitive consequences. At low enough electron density such
that only states in the lowest spin-split subbands are occupied,
states near the Fermi energy $\varepsilon_{\text{F}}$ will be
localized near four Fermi points. When electron density is not too
low, their spins are approximately quantized in $x$ direction. As
pointed out above, spins of states near Fermi points for
right-movers are approximately spin-down, opposite to the spin
direction of left-moving states near $\varepsilon_{\text{F}}$.
Assuming it to be possible to selectively raise (lower) the
electrochemical potential of right-movers (left-movers), a
spin-polarized current could be generated. Usually, creating a
population of left-movers and right-movers with different
electrochemical potentials is achieved by coupling the quantum
wire adiabatically to ideal contacts\cite{landauer2,butt:ibm:88}.
However, the underlying assumption that excess electrons injected
from the right (left) reservoir will only be spin-up (spin-down)
is not realistic because, typically, RSOC will be absent in the
contacts. The different nature of electron states in the wire and
the leads will result in strong scattering at wire-lead
interfaces. Similar to the approach taken in
Ref.~\cite{mire:prb:01}, we model this situation by attaching
semi-infinite leads with $k_{\text{so}}=0$ to the wire where
$k_{\text{so}}\ne 0$. The transmission problem can be solved
exactly by matching appropriate {\it ans\"atze\/} for wave
functions in the wire and the leads. The usual condition for
ensuring current conservation has to be modified because the group
velocity for electrons in the quantum wire with RSOC
reads\cite{mol:prb-rc:01,uz:prl:02}
\begin{equation}\label{velop}
v_y = \hbar (k_y + k_{\text{so}}\sigma_x)/m \quad .
\end{equation}
Despite the unusual spin structure at
the four Fermi points which is asymmetric with respect to
right-movers and left-movers, no spin-polarized current is
generated {\em in the leads}. However, as is shown in
Fig.~\ref{convert}, a process of current conversion occurs close
to the interfaces in the wire that results in a {\em finite spin
polarization of current in the wire}. We have therefore found a
unique type of spin accumulation that is not, as in the usual
case\cite{son:prl:87}, induced by ferromagnetic contacts. Our
analysis shows that current conversion is enabled by scattering
into evanescent modes of the wire because of the peculiar form of
the velocity operator~(\ref{velop}). A four--terminal measurement
with ferromagnetic contacts as weakly coupled voltage probes
should enable experimental verification of spin accumulation in
the wire.

Finally, we briefly remark on the effective low-energy description
of an interacting quantum wire with strong RSOC. In the spirit of
Tomonaga-Luttinger models\cite{tom:prog:50,lutt:jmp:63} for
interacting 1D systems, we linearize the single-electron energy
spectrum close to the four Fermi points. We explicitly avoid
attaching any spin labels. Rather, we define type-A (type B)
right-movers and left-movers having {\em the same\/} velocity
$v_{\text{A}}$ ($v_{\text{B}}$). Typical electron-electron
interactions give rise to a term $H_{\text{int}}= \frac{1}{2}
\int_{x,y\atop x^\prime, y^\prime}\psi^\dagger\psi(x,y) U(x-
x^\prime,y-y^\prime)\psi^\dagger\psi(x^\prime ,y^\prime)$
in the electron Hamiltonian. In the low-energy, long-wave-length
limit, we can write $\psi(x,y)=\sum_{\alpha=\text{A,B}\atop\beta=
\text{R,L}}\psi_{\alpha\beta}(y)\Phi_{k_{\text{F}\alpha\beta}}(x)$
and assume $U$ to be long-range on the scale of the wire width but
short-range on the scale of the wire length. It is important to
note that the present case differs from the usual one in that the
transverse wave-function spinors $\Phi_{k_{\text{F}\alpha\beta}}
(x)$ are {\em nearly orthogonal}. As a result, backscattering
processes are strongly suppressed. Apart from this fact and the
peculiar spin structure of states near the four Fermi points, the
present system is identical, on a formal level, to a
two-component\cite{penc:prb:93} or Zeeman-split\cite{aoki:prb:96}
Tomonaga-Luttinger model. Response to an external magnetic field
will, however, be special in the present case. Postponing a
detailed analysis to a later publication, we mention here only a
few basic facts. When Fermi points
are far enough away from anticrossings, a magnetic field $B$
applied in $x$ direction will shift right-movers (left-movers) to
higher (lower) energies. [See Fig.~\ref{avspin}b).] The Zeeman
term in bosonized form reads then $H_{\text{Z}} = -\frac
{\Delta_{\text{Z}}}{\sqrt{2\pi}} \int_x\Pi_\rho$, where $\Pi_\rho$
is canonically conjugate to the phase field $\theta_\rho$ that is
related, within the usual\cite{voit:reprog:94} phase-field
formalism, to the total electron density $\rho_{\text{tot}}=
\sum_{\alpha=\text{A,B}\atop\beta=\text{R,L}}\rho_{\alpha\beta}$
via $\sqrt{2/\pi}\partial_y\theta(y)=\rho_{\text{tot}}(y)$.
Approximate orthogonality of transverse parts of electron wave
functions enables spin-flip processes, in the long-wave-length
limit, only between left-moving and right-moving branches of the
same type (A or B). In general, any spin-flip process incurs a
large momentum transfer.

This work was supported by the DFG Center for Functional
Nanostructures at the University of Karlsruhe.

\end{document}